\documentclass[aps,pra,superscriptaddress,twocolumn,10pt]{revtex4-1}
\usepackage{graphicx}
\usepackage{bm}
\usepackage[usenames,dvipsnames,svgnames,table]{xcolor}
\usepackage{epsfig}
\usepackage{amsmath}

\usepackage{amssymb}
\usepackage{array}
\usepackage{nccmath}
\usepackage{dsfont}
\usepackage{array}

\begin{document}
	\title{Front dynamics in the Harper model}
	\author{Gerg\H{o} Ro\'osz$^1$}
	\date{$^1$Wigner Research Center for Physics, Budapest
		\today}
	
	\begin{abstract}
		The front dynamics in the Harper (or Aubry-Andr\'e) model (which has a localization transition) is investigated using two different settings, particle number front where the system is at zero temperature, and initially, the particle numbers differ on the two sides, and temperature front where the two sides have different temperature initially. The two differently prepared half systems are connected suddenly, and the following dynamics is investigated.
		In the extended phase, the dynamics is ballistic, similar to the dynamics of a pure system. At the critical point, one finds a power-law time dependence of the particle number and the entanglement entropy of the zero temperature setting. In the localized phases, the observables oscillate around an average value, which is independent of the system size. 
		The particle number front shapes have been investigated at zero temperature setting, in the extended phase they scale together exactly as in the homogeneous XX chain, however at the critical point the scaling relation contains a power ($t^{0.55}$) of the time.
		The mutual information between neighboring intervals at the front has been calculated, and it is promotional with the logarithm of the interval length and also with the logarithm of the time in the extended phase and at the critical point. The prefactors of the time and size dependence are equal for the zero temperature process, however differ for the finite temperature front.   
	\end{abstract}
	\maketitle
	
	When two systems prepared in states with different properties (e.g. particle number, temperature) are suddenly connected a nontrivial dynamical process starts, where the disturbed region (so-called front) broadens with time starting from the point of connection, this process is usually referred as front dynamics.

	From analytical results about the XX chain \cite{antal1999, ogata2002, karevski2002, hunyadi2004} one learns, that as the disturbed region broadens with time, its shape varies. In the infinite system, the front broadens without limit, in a finite system the front is reflected from the ends.   For example, the magnetization in the XX chain when the starting state is a magnetic kink evolves according to a simple scaling function $m(n,t)=\Phi(n/t)$ ($n$ is the place coordinate, $t$ is the time). This means the ballistic broadening of the front.

	In one of the first studies about dynamical steady states \cite{antal1997, antal1998}  the authors proved that the steady state in the XX chain can be prepared as a ground state of the Hamiltonian with current generators added. 
	
	A later study \cite{viktor2014} investigated the XX chain after connecting two half-infinite segments with different temperatures (I call it here temperature-front). They investigated the mutual information between two neighboring intervals and bought intervals inside the front region when the steady state already built up. It has been found, that the mutual information grows with the log of the interval size $I \sim \ln l$, and (between the two half infinite chains) with the log of the time $I \sim \ln t$, in addition the two prefactor equals.
	There are a series of further results about the details of the XX front dynamics, the statistic of particle numbers at the front has been calculated \cite{viktor2013}, there are analytical results about the magnetization profile scaling in external magnetic field \cite{platini2007}, the spin current fluctuations have been obtained \cite{hiroki2019}, the entanglement Hamiltonian has been obtained using bosonization  \cite{federico2022}.
	In the transverse Ising, and in the XY model \cite{karevski2002, marton2017, gambassi2017, dasgupta2017, viktor2016, marton2017B, platini2006, platini2005, viktor2020}, the general scaling of the front is similar to the front of the XX  model, with several minor differences. 
	
	Several studies has investigated the XXZ chain \cite{scopa2022, lopez2021, collura2020, vir2019, braunfield2017, misguich2013, einhellinger2012, lancester2010, langer2009, antal2008, dominique2005}. From these results about the XXZ chain I would like to highlight,  that although the dynamics is ballistic (like in XX or Ising chains) for most initial states, there are special initial states with sub-diffusive behavior \cite{vir2019}, and a region has been found, where the transport stops \cite{dominique2005}.  So the dynamics is strongly dependent on the initial state, which is an important difference compared to quench dynamics, where the effect of the initial state is small \cite{roosz2014}. 
	
	To talk about methods, there are results in exactly solvable models, for example, a canonical transformation \cite{lieb61} after a Jordan-Wigner transformation \cite{jordan1928} in the XX and XY and Ising chain,  and using other analytical solutions in the harmonic chain \cite{zimbi2014}, or the  Sine-Gordon \cite{XHorvath2022} model. 
	When the analytical solution is not known, one may approximate the original model with a solvable one, (in one dimension it is often a Luttinger liquid approximation obtained by bosonization \cite{lancester2010,federico2022,collura2020,braunfield2017}) or follow the time evolution with a numerical method, such as time-evolving block decimation \cite{misguich2013, einhellinger2012} or time-dependent DMRG \cite{langer2009, dominique2005}.
	
	On the other hand, one may concentrate on the physical impact of local disturbances rather than strictly following the unitary dynamics. Local disturbances in one-dimensional quantum systems create an effect, which spreads according to a light cone.      This physical observation led to the invention of the quasi-classical description of the dynamics of one-dimensional non-interacting quantum systems, first applied to the description of global quenches \cite{calabrese2007, calabrese2005, igloi2011, rieger2011}.
	This method and its generalization to Bethe ansatz integrable systems has been  successfully applied to the front dynamics, \cite{alba2019, bertini2018, castro2016, ruggiero2020, doyon2020, misguich2019, alba2021, dubail2022, fagotti2017, ruggiero2021, bastellino2019, bulchandani2017, doyon2018, schemmer2019, malvania2021}
	Depending on the model, precise approximations or even exact results have been obtained using semi-classical dynamics and generalized hydrodynamics. 
	
	For non-interacting aperiodic and disordered models, numerical results for global sudden quenches have been qualitatively interpreted assuming that signals spread with anomalous diffusion (aperiodic systems) or with the logarithm of the time (in disordered systems) \cite{igloi2012, igloi2013, roosz2017}. 
	
	The goal of the present study is to investigate the front dynamics at a localization transition and answer the questions if the shape of the fort is different at the transition point, what is the behaviour of the mutual information in the front region?
	
	The rest of the paper is organized as follows. In section I. the model is defined and its equilibrium properties are presented. In section II. the local particle numbers, entanglement entropy, and mutual information are defined. In section III. the numerical results are presented, in section IV. I discuss the results. In the Appendix technical details of the time evolution are given.
	
	\begin{figure}
		\includegraphics[width=0.9\linewidth]{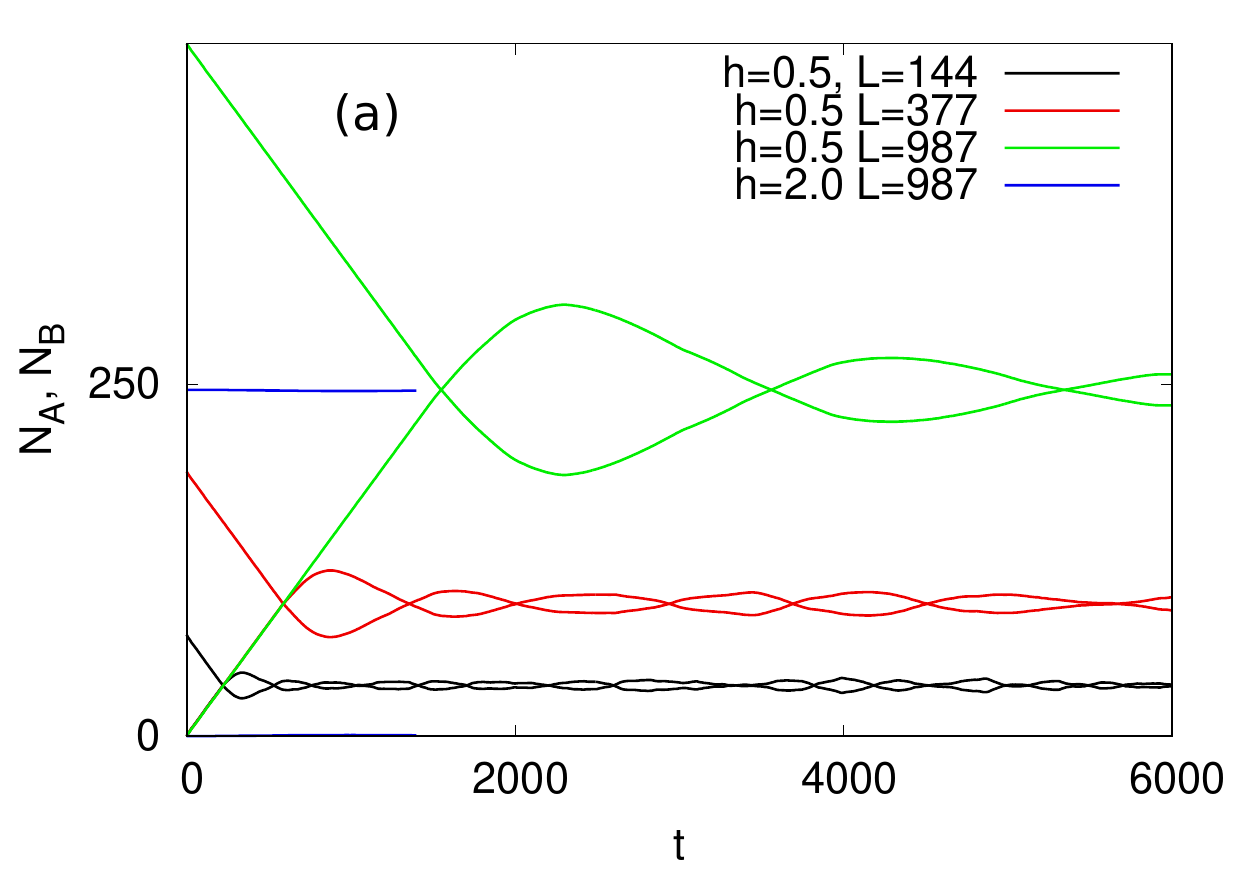}
		\includegraphics[width=0.9\linewidth]{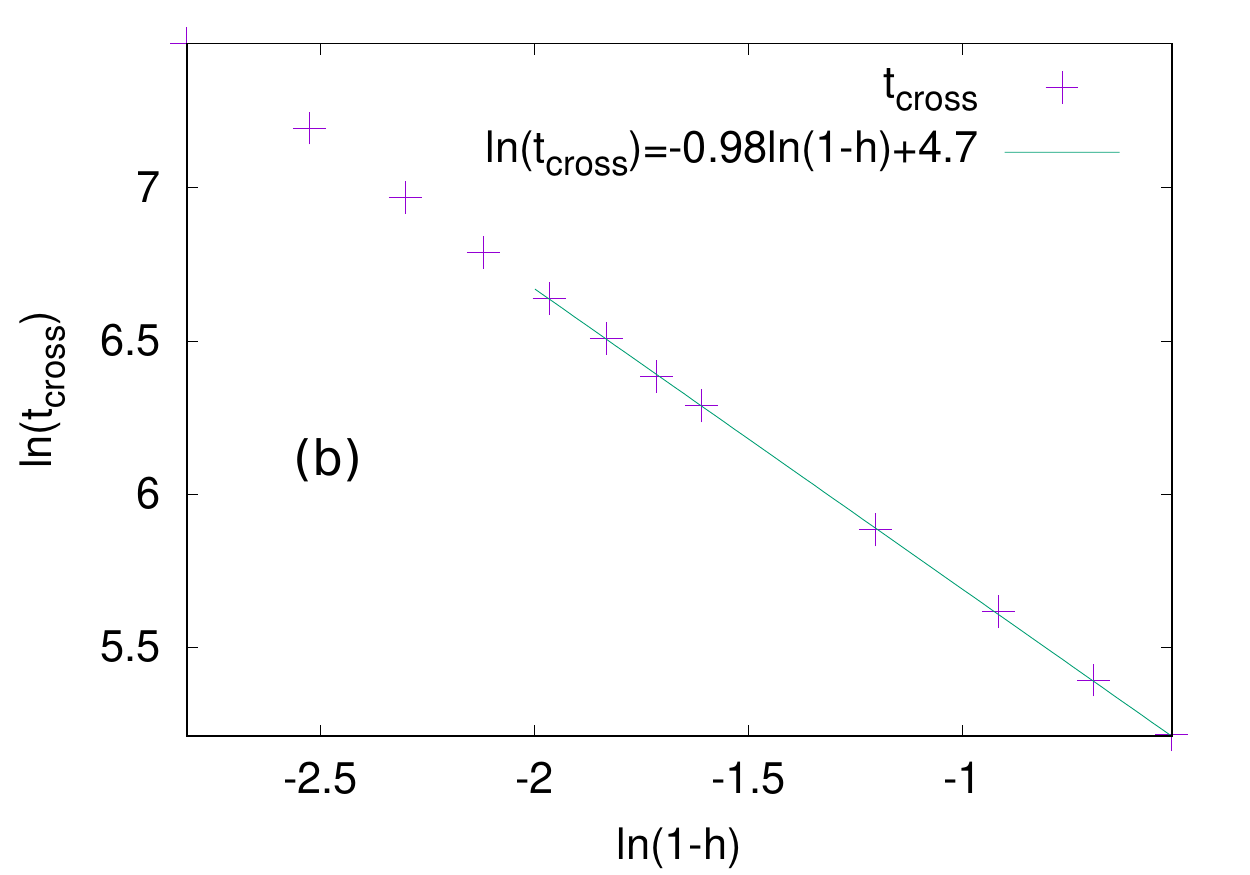}
		\caption{Typical behavior of the particle numbers of the two subsystems at the zero temperature protocol. The colors denote the different quench parameters ($h$ and $L$). The particle numbers $N_A$ and $N_B$ from the same process are denoted by the same line type, one grows initially, and its pair decreases initially. \label{fig:tip_vis}
			The logarithm of first crossing times of the particle numbers, as the function of $\ln(1-h)$, for  $L=144$.    \label{fig:crossingtimes}	} 
	\end{figure}
	
	\begin{figure}
		\includegraphics[width=0.9\linewidth]{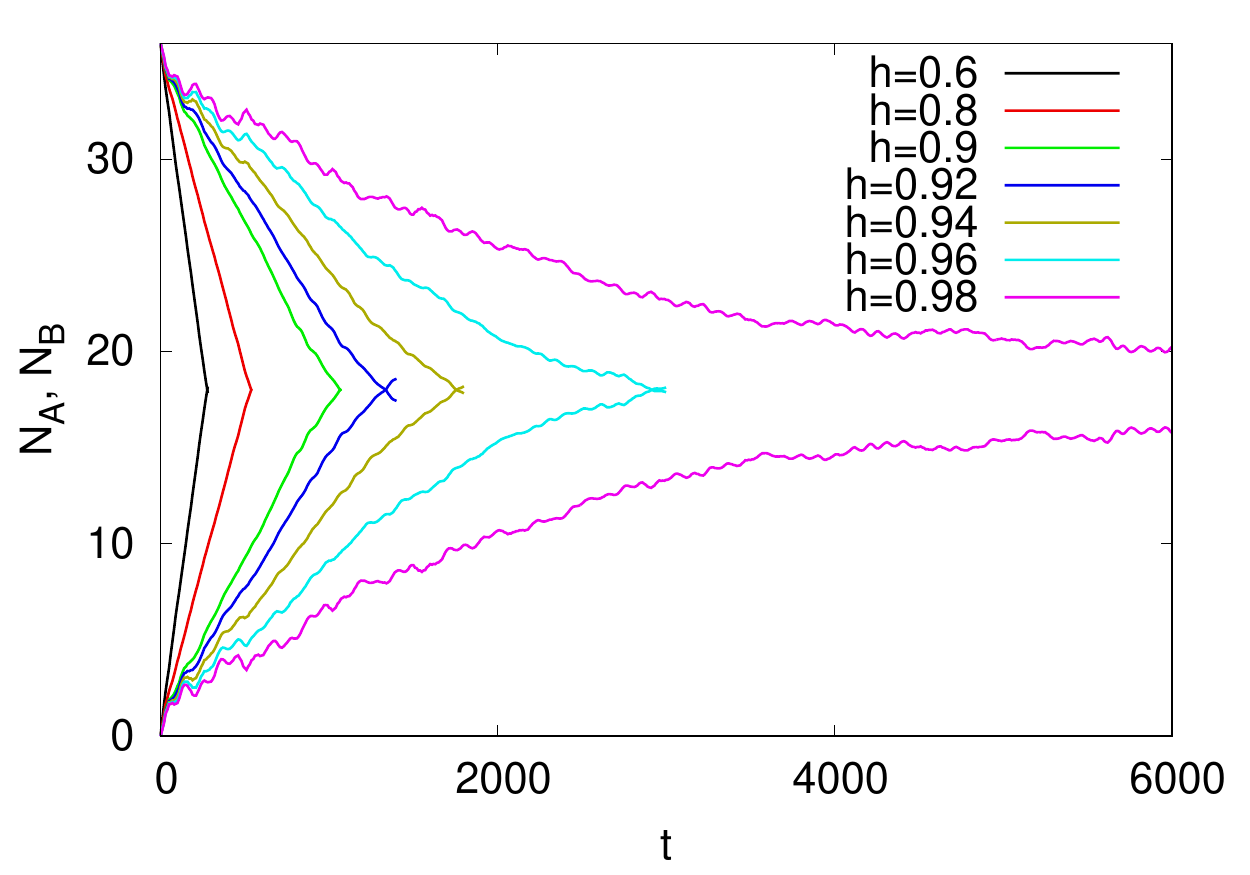}
		\caption{Particle numbers of the two halves at different after quench $h$ fields, for the same system size $L=144$, zero temperature protocol. For oscillating parts, the curves are not shown, just the first crossing points. \label{fig:Nvsh}  }
	\end{figure}
	
	\begin{figure}
		\includegraphics[width=0.9\linewidth]{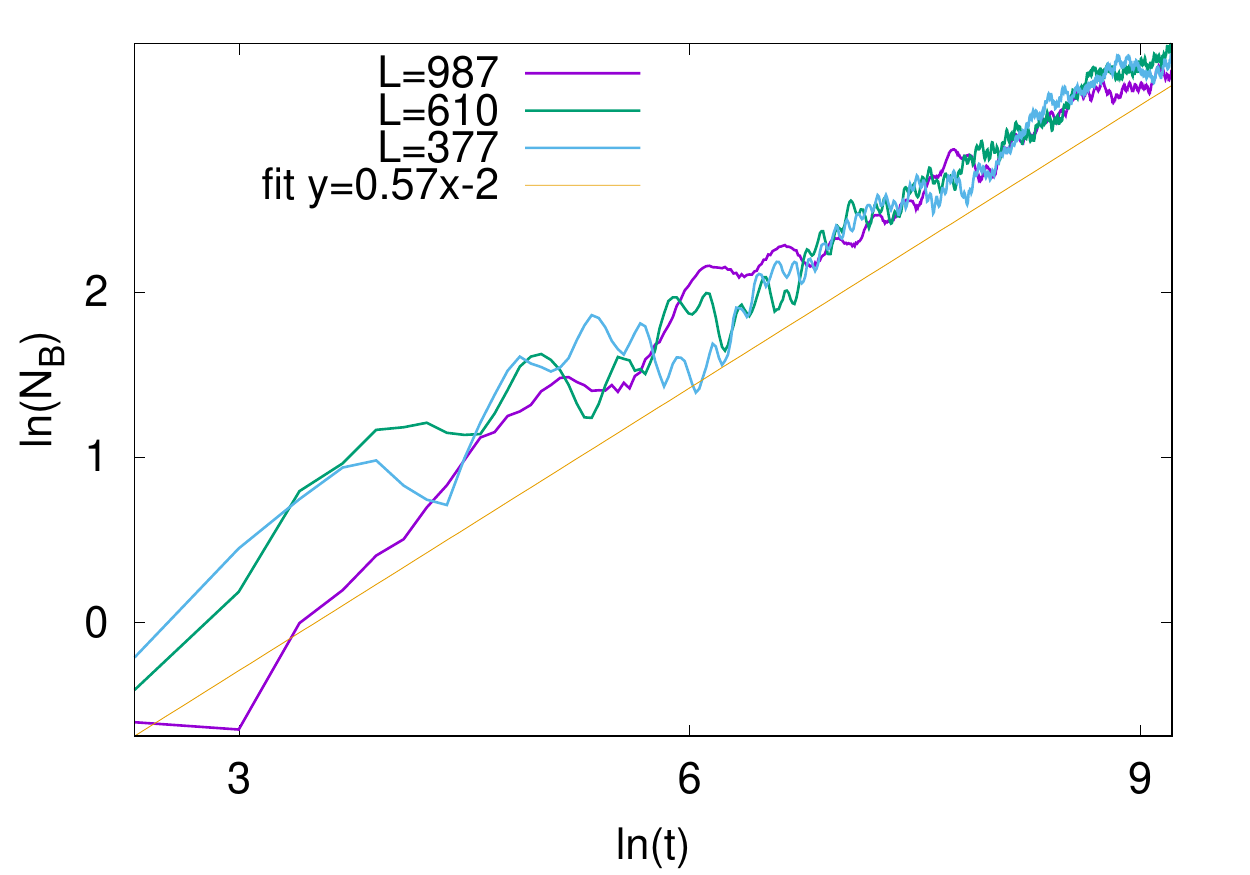}
		\caption{Variation of the particle number (in the initially empty half) at a critical quench ($h=1$) for various system sizes, for the zero temperature protocol. The straight line is a fit. \label{fig:NCRIT}}	
	\end{figure}
	
	\begin{figure}
		\includegraphics[width=0.9\linewidth]{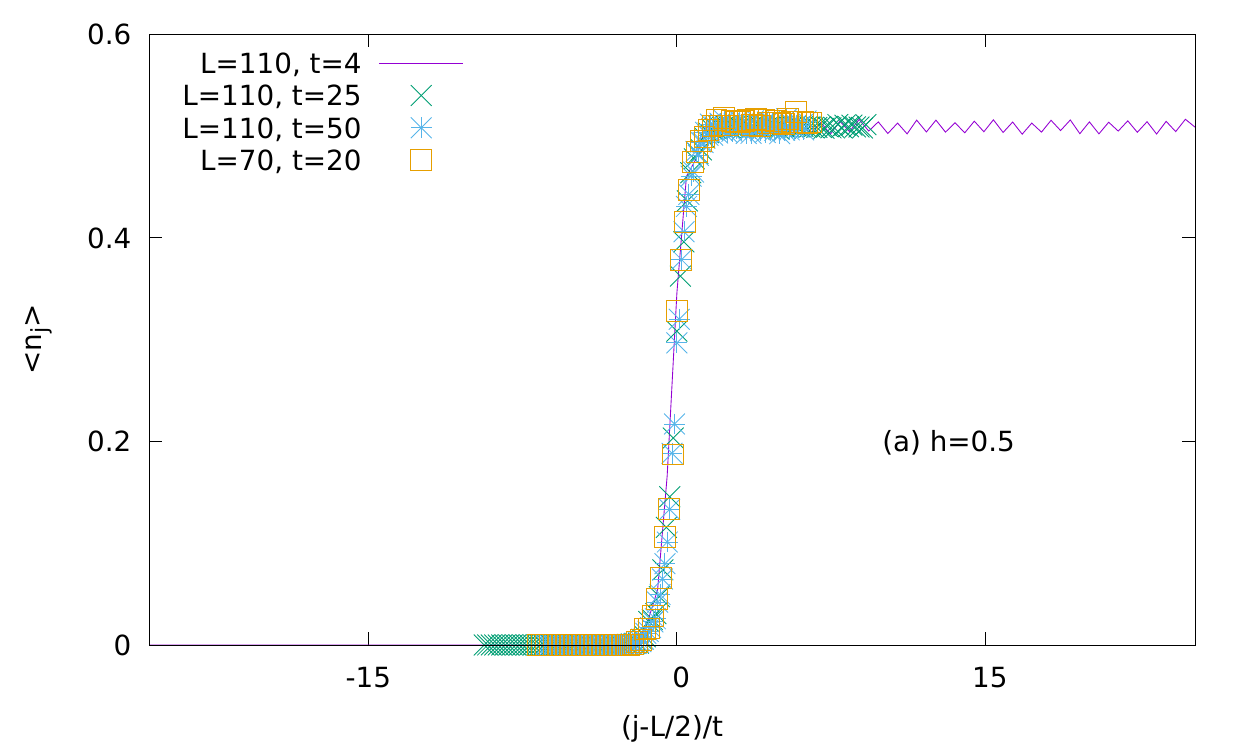} 
		\includegraphics[width=0.9\linewidth]{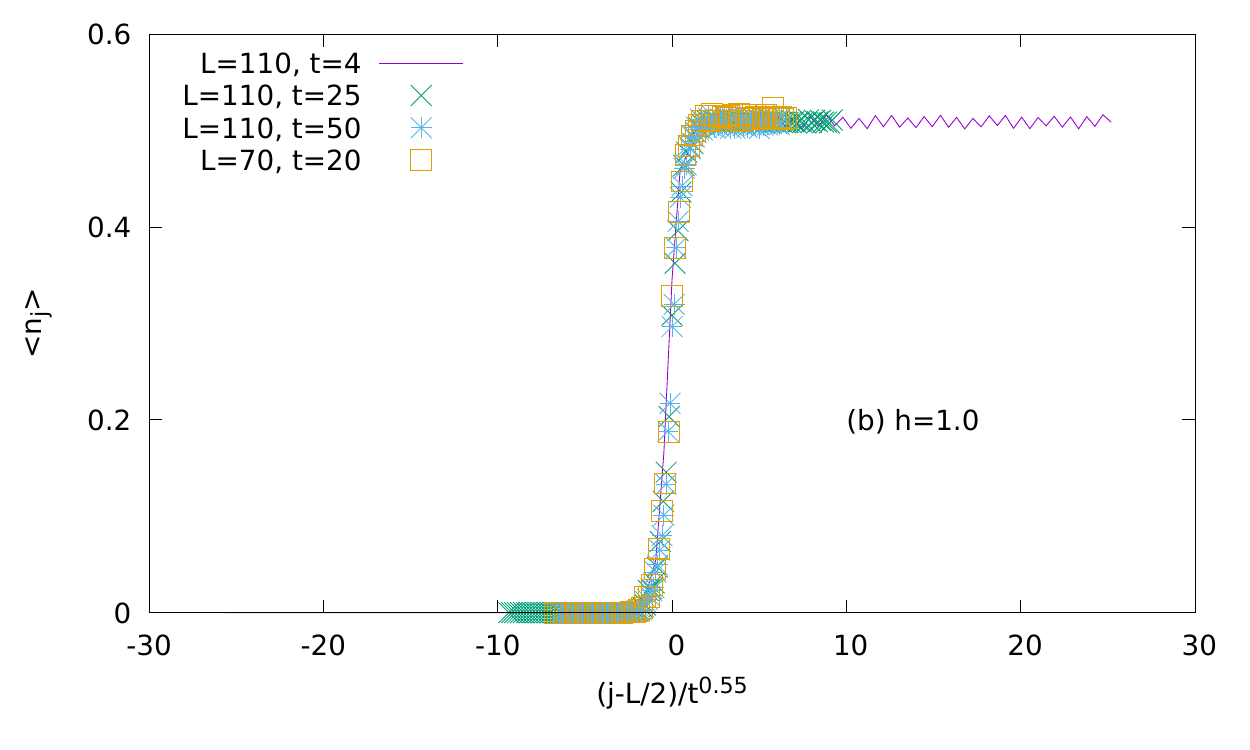}
		\includegraphics[width=0.9\linewidth]{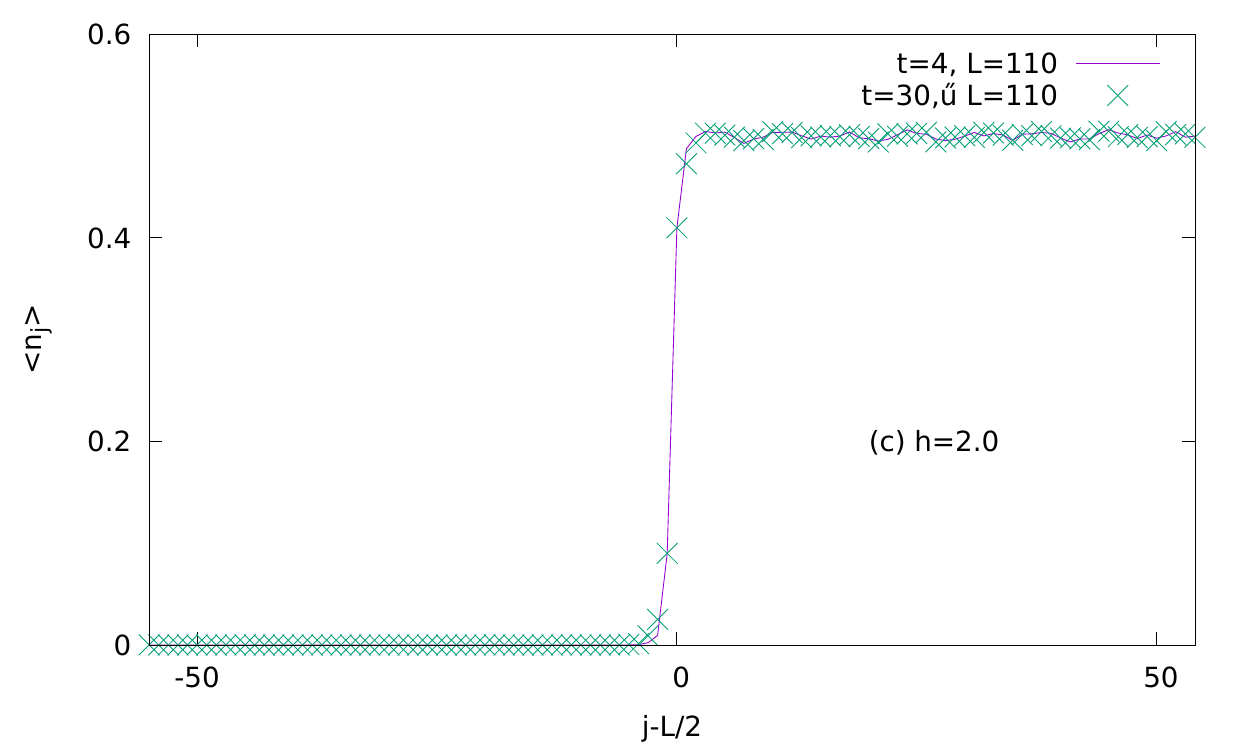} 	 
		\caption{Shape of the particle number front.  The  expectation value of the particle number $n_j$ at position $j$  in the extended phase (a), the critical point (b) and the localized phase (c), as the function of $(j-L/2)/t^\alpha$  for various times and for $L=50,70,1110$ \label{fig:scaling}}
	\end{figure}
	
	\section{Model}
	The Harper model is defined as follows
	\begin{equation}
		H=-\frac{1}{2}\sum_{l=1}^{L-1} c^{\dagger}_l c_{l+1} + c^{\dagger}_{l+1} c_l  + h \sum_{l=1}^L \cos\left( 2 \pi \kappa l \right) c^{\dagger}_l c_l
		\label{eq:ham}
	\end{equation}
	where $c_l$ $c^{\dagger}_l$ are the fermionic annihilation and creation operators for $l=1 \dots L$, $L$ is the length of the systems, $h$ is an external parameter (the amplitude of the inhomogeneity), and $\kappa=(1+\sqrt{5})/2$.
	There is a localization transition in the model \cite{aubry1980}, for $h<1$ every eigenstate are delocalized, free-wave like, for $h>1$ every eigenstate is localized. The localization length for $h>1$ is given by \cite{aubry1980, thouless1983} 
	\begin{equation}
		h_{loc}=\frac{1}{\ln h}
		\label{eq:loc_lnegth}
	\end{equation}
	The system is self-dual to the critical point: (a modified) Fourier transform maps the $H$ operator with $h$ to a similar operator with $1/h$ \cite{aubry1980,harper1955,roosz2014, thouless1983}.
	The localization transition occurs for every irrational $\kappa$,  the localization length (E.q. (\ref{eq:loc_lnegth})), and the self-duality are also not sensitive to $\kappa$, until $\kappa$ is irrational \cite{evers2008}. However, the details of the transition are sensitive to the $\kappa$ value. It has been shown with renormalization group studies \cite{tessa2020} and numerical calculations \cite{sinha2019}, that the $z$ critical exponent depends on $\kappa$.

	The spectrum is continuous in the extended phase, fractal at the critical point, and pure point spectra in the localized phase \cite{wilkinson1994}.
	The one-particle eigenstates are multifractals at the critical point \cite{evang200,siebesma87}. 
	In the ground state at half filling  \cite{roosz2020} the entanglement entropy  of an interval of length $l$ scales as 
	\begin{eqnarray}
		S= 0.33 \ln l
	\end{eqnarray}
	so the scaling is identical with the homogeneous XX chain ($c_{\textrm{eff}}=1$). At the critical point of the system, the effective central charge changes \cite{roosz2020}
	\begin{equation}
		S=0.21 \ln l \;,
	\end{equation}  
	this corresponds to $c_{\textrm{eff}}\approx 0.78$
	In the localized phase the entanglement entropy saturates to a constant value in the $L\to \infty$, $l\to \infty$ $l/L=$const.  limit. This entropy depends on the localization length as
	\begin{equation}
		S_{\textrm{sat}} \sim  \frac{c_\textrm{eff}}{3} \ln (l_{\textrm{loc}}) \;.
	\end{equation}  
	The scaling of the logarithmic negativity follows the same effective central charges \cite{roosz2020}.
	
	There are gaps in the spectrum of the Aubry-Andre model, bought in the extended phase and in the critical points \cite{aubry1980} \cite{roy2019}. If one chooses the chemical potential (Fermi surface) at the location of one of these gaps, the system becomes non-critical (gapped), and the entanglement entropy follows the area law, which in one dimension means it remains constant \cite{roy2019}. In the present work, zero chemical potential has been used during the time evolution. 
	
	After a quantum quench (sudden change of $h$) the dynamics depends mainly on the after-quench Hamiltonian \cite{roosz2014}. If the after-quench Hamiltonian is in the extended phase the dynamics resemble to the dynamics of a homogeneous XX chain, and the entanglement entropy grows linearly in time $S\sim t$ \cite{roosz2014}. At the critical point, the entanglement entropy grows with a power function of the time $S \sim t^{\sigma}$, and in the localized phase it remains bounded.  
	
	The model has been realized in an optical lattice with cold atoms \cite{hauschild2015, rozenheimer2013,modungo2009,deisler2011} and step-like initial condition of occupation number has also been realized in cold atom experiments \cite{choi2016}, so our zero temperature particle number protocol might be possible to realize in these experiments.
	\section{Quantitates of interest}
	I divide the system into two halves $A$ and $B$, where $A$ contains the first $[L/2]$ sites (where $[\dot]$ is the lower integer part) and $B$ contains the rest of the system.  The two halves are not connected for $t<0$  and are prepared in states with different physical parameters.
	At $t=0$ the two halves of the system are connected, and a non-trivial time evolution starts. 
	
	For the zero temperature protocol, the particle number is different at $t=0$ in the two halves of the system, the $A$ subsystem is empty, and the $B$ subsystem is half-filled.
	
	In this case the system at $t=0$ is in a pure state $|\psi_0\rangle$, and the dynamics is given by the Schr\"odinger equation $i \frac{\partial}{\partial t}|\psi(T) \rangle = H |\psi (t) \rangle  $ 
	
	For the finite temperature protocol, at $t=0$ the $A$ subsystem is at thermal equilibrium at $T_A$ temperature, and the $B$ subsystem is at thermal equilibrium at $T_B \neq T_A$ temperature, and the two halves are not connected. We connect $A$ and $B$ at  $t=0$, and the dynamics for $t>0$ is governed by the Hamiltonian e.q. (\ref{eq:ham}). In our model, the couplings to the heat reservoirs which have created the initial thermal equilibrium (at two different temperatures) of the two halves are not present,  and the dynamics is given by the Schr\"odinger equation. This corresponds (in experiments) to short time scales compared to the thermal equilibration timescale.

	In this second protocol, the left/right subsystems are prepared in the
	\begin{align}
		\rho_A & = \frac{1}{Z_A} \exp[-\beta_A H_A] \\
		\rho_B & = \frac{1}{Z_B} \exp[-\beta_B H_B] 
	\end{align} 
	states, and the whole system is prepared in the
	\begin{equation}
		\rho=\rho_A \otimes \rho_B
	\end{equation} 
	state. For $t>0$ the dynamics is given by the $i \frac{\partial}{\partial t} \rho(t) = [H, \rho]$ equation. 
	Technically, I follow the time evolution of the operators in the Heisenberg picture, so the dynamics of the two protocol is common, only the initial state differs.
	
	I investigate (for both protocols) the time evolution of the particle numbers in the two sectors ($N_A$, $N_B$), and for the zero temperature protocol the entanglement entropy of the two halves, and for the finite temperature protocol the mutual information between the two halves. 
	For the zero temperature protocol, I also investigate the finite size scaling of the mutual information in the dynamical steady state \cite{zimbi2014}. This means the mutual information is calculated between two intervals (of length $l$) at the border between the two subsystems for interval lengths sorter than the half system size for a given time $t>0$. In this setting, one interval is from the $L/2-l$th site to the $L/2$th site, and the other interval is from the $L/2+1$th site to the $L/2+l$th site.  
	
	Let's take a closer look at the aforementioned quantities. The particle numbers are defined as the expectation values of the particle number operators
	\begin{align}
		\hat{N}_A &= \sum_{l \in A} c^{\dagger}_l c_l \\
		\hat{N}_B &= \sum_{l \in B} c^{\dagger}_l c_l
	\end{align} 
	and $N_A=\langle \hat{N}_A \rangle$ and $N_B=\langle \hat{N}_B \rangle$. 
	
	To define the mutual information and the entanglement entropy \cite{bennett1996}, one defines the reduced density matrices of the $A$ and $B$ subsystems. The density matrix of the whole system is $\rho(t)$, this include the special case of the pure state $|\psi(t) \rangle$ as a projector $\rho_{\textrm{pure}}(t) = |\psi(t) \rangle \langle \psi(t) | $.
	The reduced-density matrices are
	\begin{align}
		\rho_A &= \textrm{Tr}_B \rho \\
		\rho_B &= \textrm{Tr}_A \rho 
	\end{align} 
	Now, one defines tree entropies, the entropy of the whole system, and the entropies of the reduced density matrices.
	\begin{align}
		S &= -  \textrm{Tr} \rho \ln \rho \\
		S_A &= -  \textrm{Tr}_A \rho_A \ln \rho_A \\
		S_B &= -  \textrm{Tr}_B \rho_B \ln \rho_B
	\end{align}
	for pure states $S=0$ and $S_A=S_B$. In this case later quantity is the entanglement entropy
	\begin{equation}
		S_{\textrm{entanglement}}=S_A=S_B \;.
	\end{equation}
	For non-pure states, the entropies of the two reduced density matrices are different ($S_A \neq S_B$). In this case, one can quantify the total correlations between the two subsystems using the mutual information which is defined as follows \cite{nielsen2020}
	\begin{eqnarray}
		I = S_A + S_B - S \;.
	\end{eqnarray}

	\begin{figure}
		\includegraphics[width=0.9\linewidth]{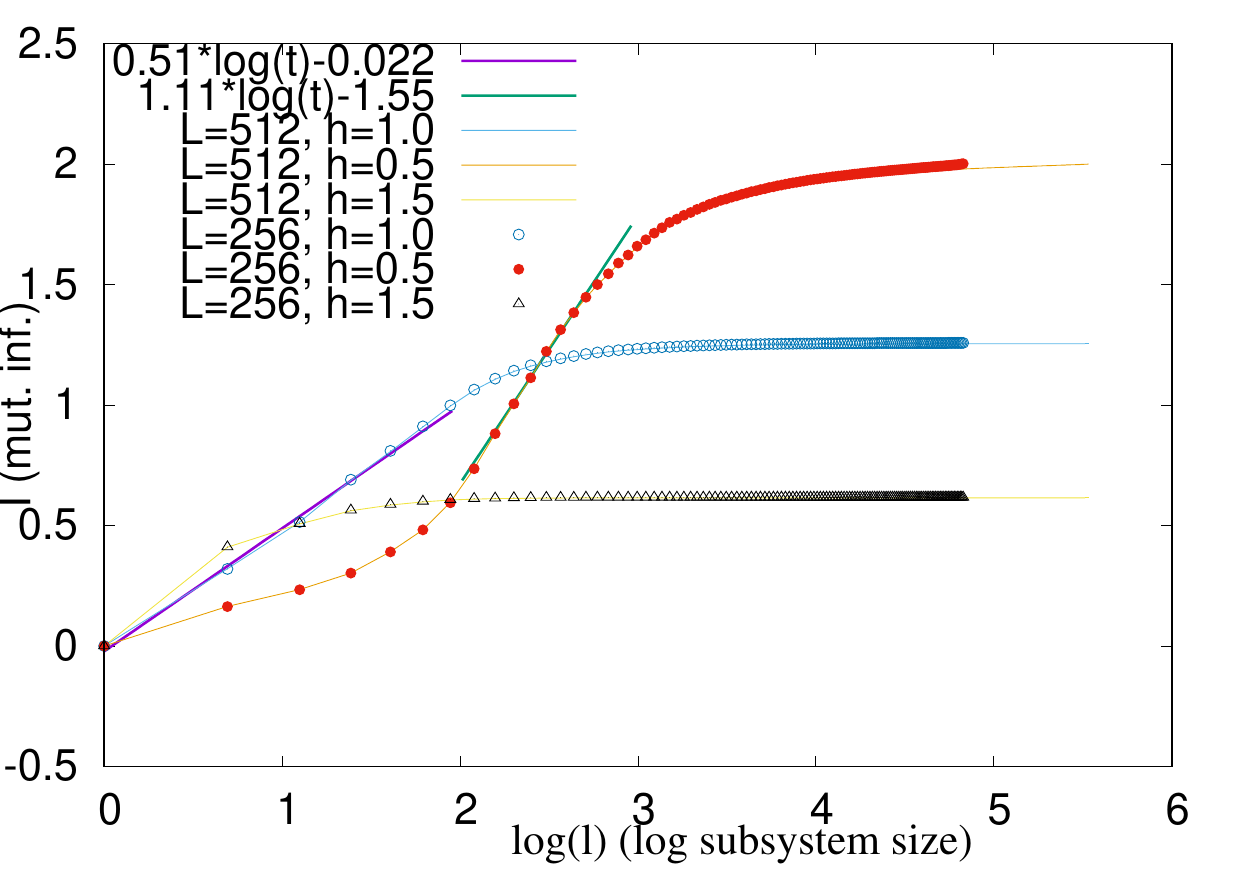}
		\caption{Mutual information after the zero temperature quench as the function of the logarithm of the subsystem size, at a given time $t=20$.  The straight lines show fit. \label{fig:zero-T-MUTI-size-dep-log}	}	
	\end{figure}
	\begin{figure}
		\includegraphics[width=0.9\linewidth]{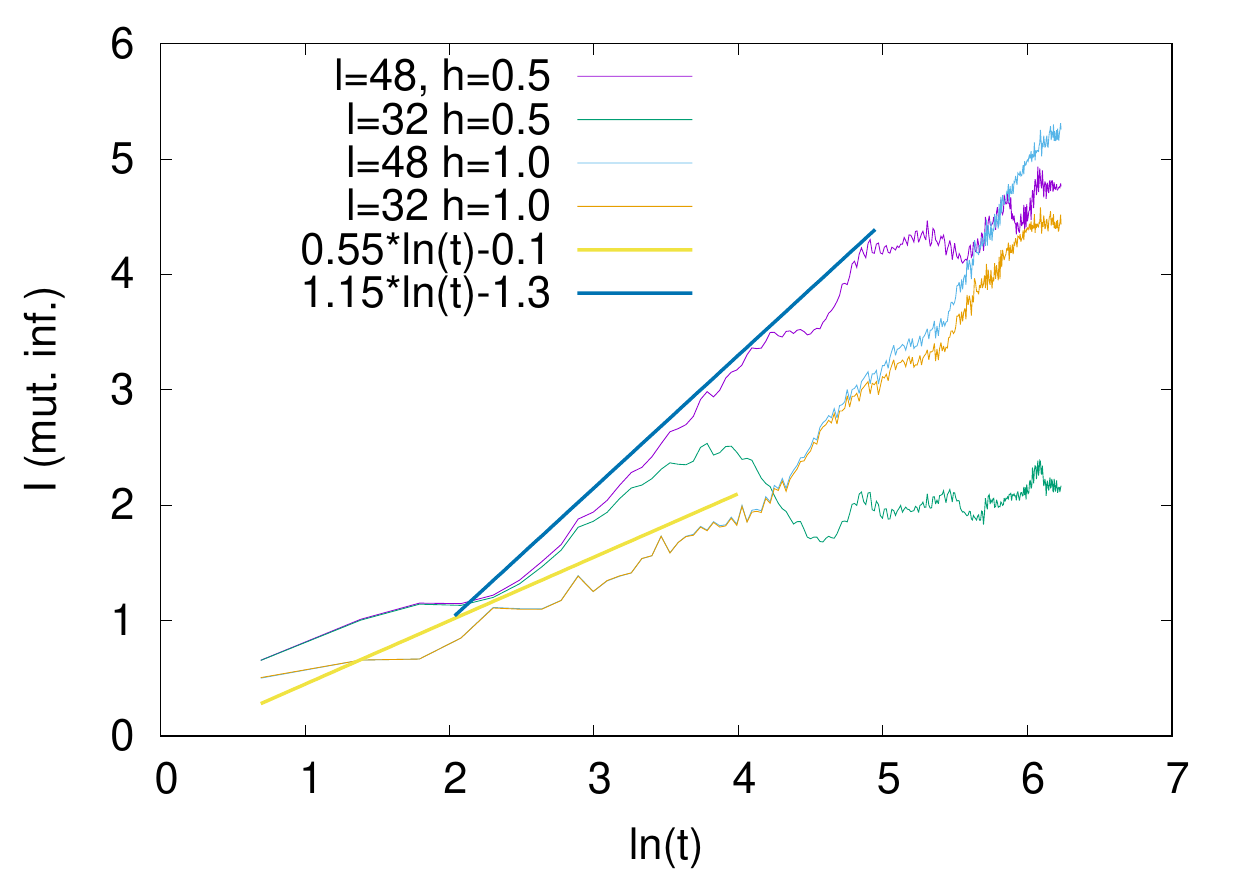}
		\caption{Mutual information after the zero temperature quench for a few fixed subsystem sizes,  as the function of the time.  \label{fig:zero-T-MUTI-time-dep}	}	
	\end{figure}
	\begin{figure}
		\includegraphics[width=0.9\linewidth]{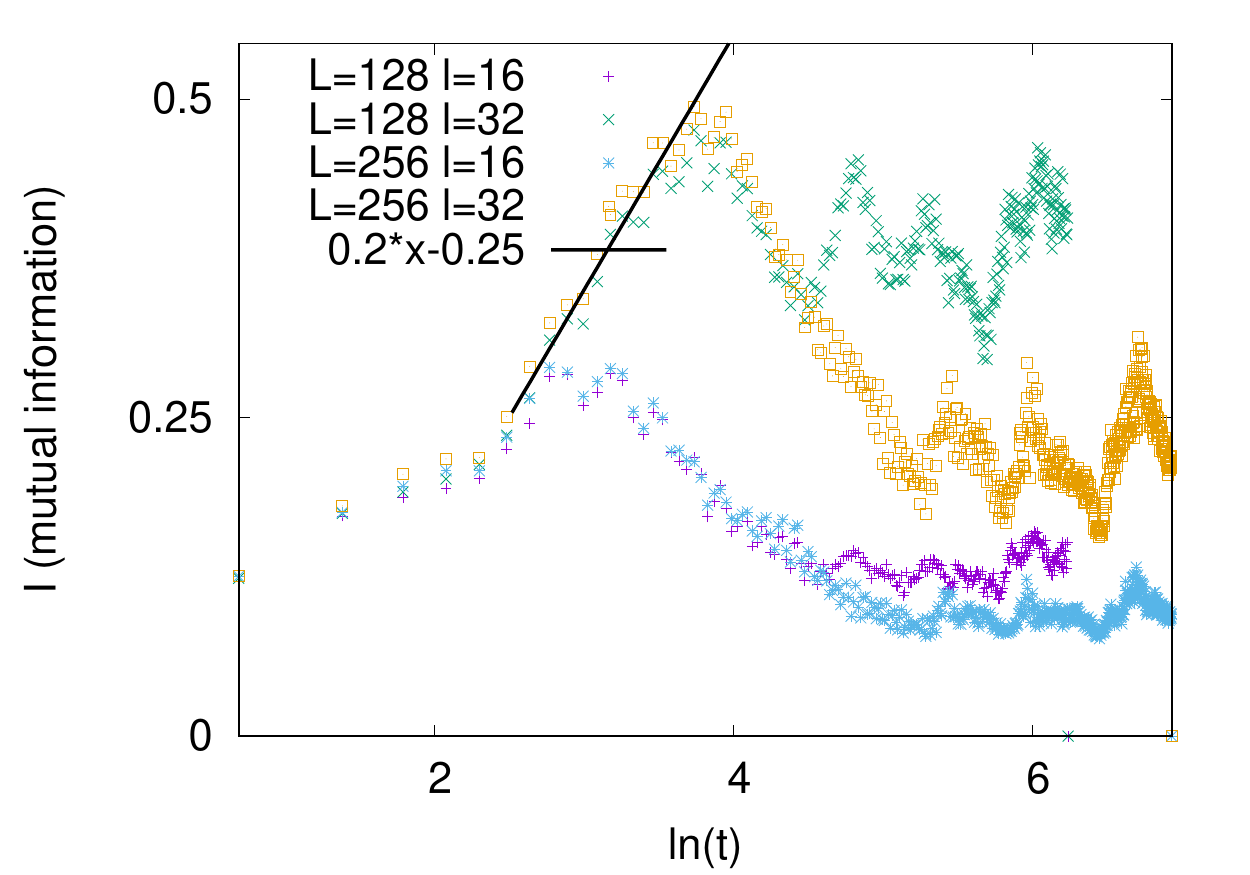}
		\caption{Mutual information after finite temperature quench for a few system and subsystem sizes, as the function of the time, in the extended phase $h=0.5$ . The data is averaged over $10^4$ randomly chosen phases.   \label{fig:fin-T-MUTI-time-dep-h0.5}}
	\end{figure}
	\begin{figure}
		\includegraphics[width=0.9\linewidth]{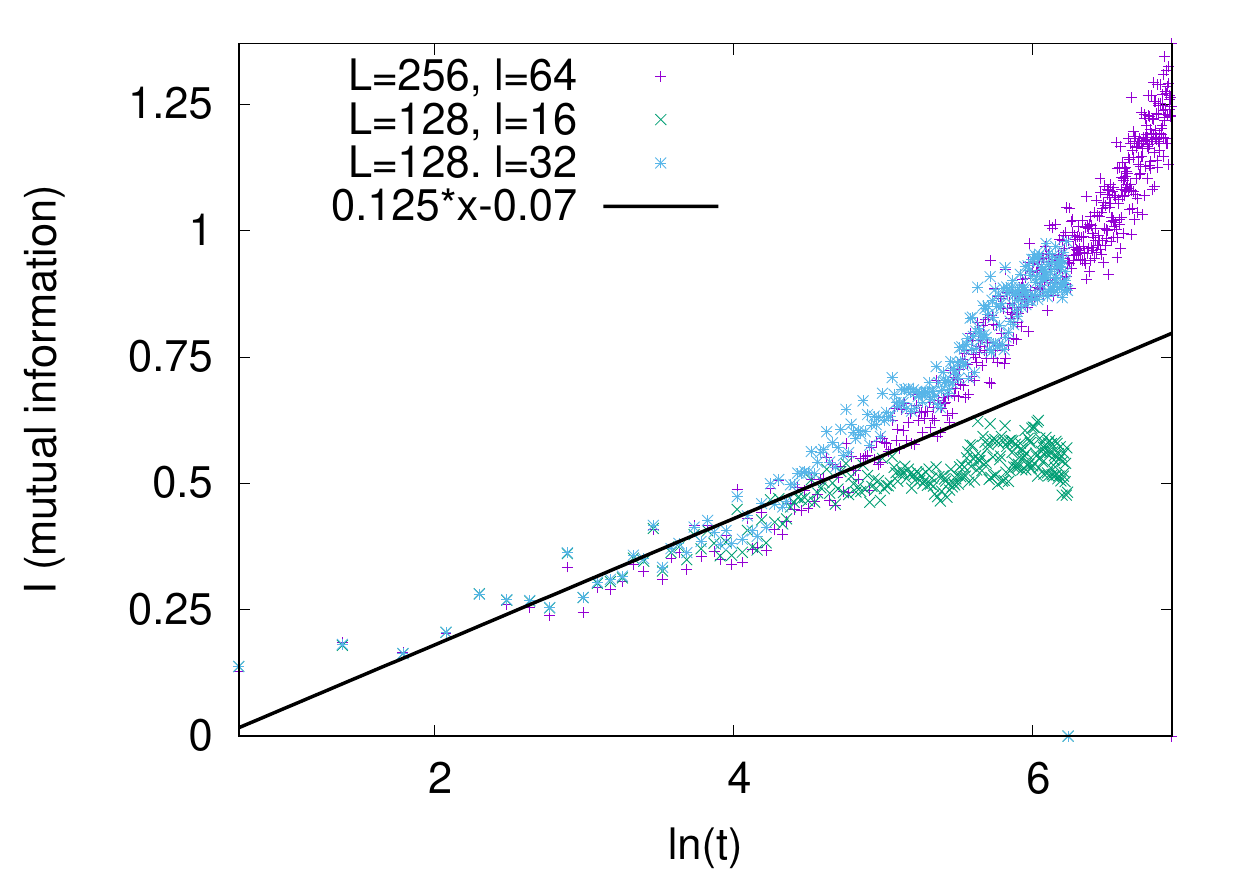}
		\caption{Mutual information after finite temperature quench for a few system and subsystem sizes, as the function of the time, at the critical point $h=1.00$ . The data is averaged over $10^4$ randomly chosen phases.   \label{fig:fin-T-MUTI-time-dep-h0.5}}
	\end{figure}
	\begin{figure}
		\includegraphics[width=0.9\linewidth]{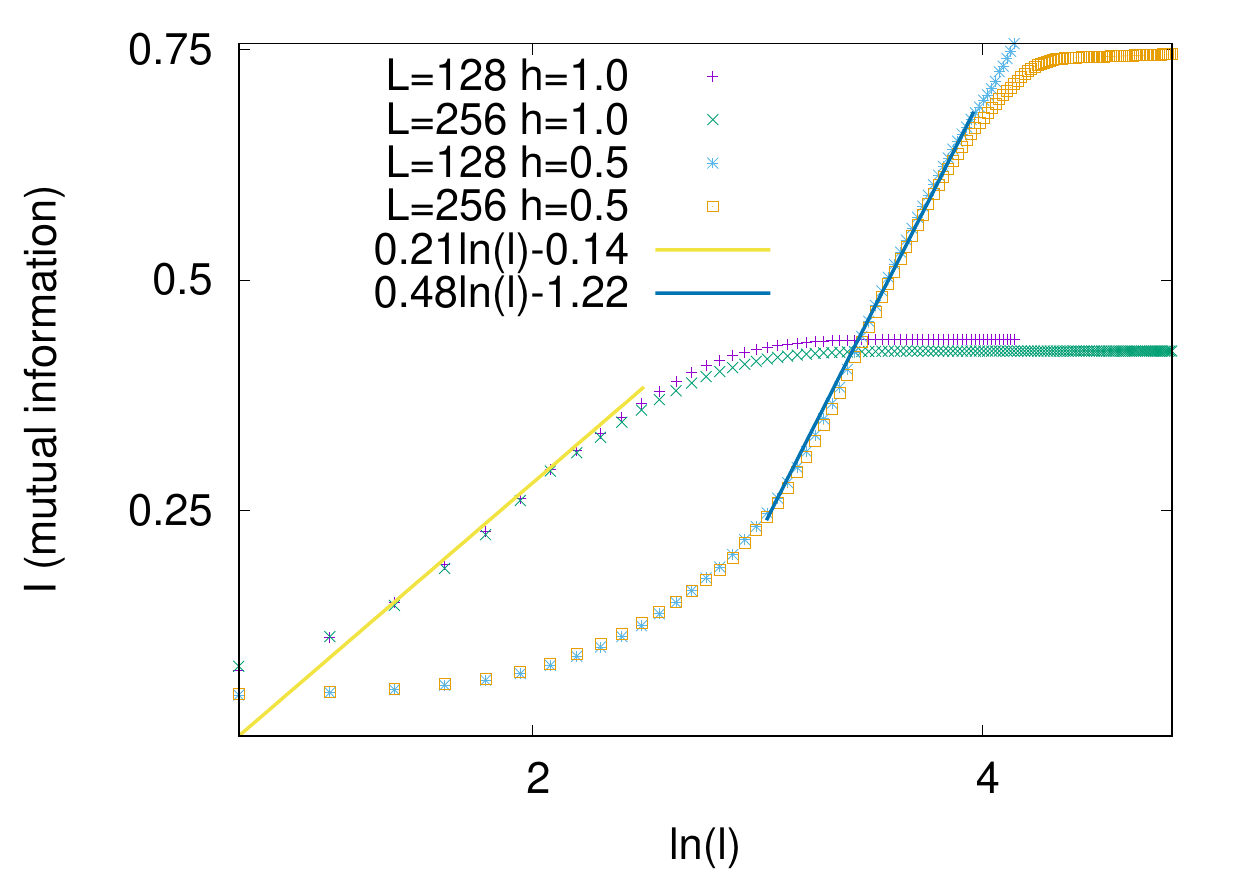}
		\caption{Mutual information after a finite temperature quench at a fixed time $t=60$, as a function of the subsystem sizes. \label{fig:fin-T-MUTI-size-dep}}
	\end{figure}
	
	\section{Results}
	In this section numerical results about the particle number quench are presented. The typical behaviour of the particle number after this quench is shown in Fig. \ref{fig:tip_vis}, in this figure, there are data about quenches in the extended phase (at $h=0.5$ ) with three different system sizes ($L=144, 377,  987$), and there is one quench to the localized phase ($h=2.0$). 
	One can see, that in the localized phase the dynamics "froze" the expectation value of $N_A$ and $N_B$ remains close to the initial values. In the extended phase, the particle numbers change. For short times the particle number of the initially empty subsystem grows linearly, and the particle number of the initially half-filled subsystem decreases linearly with the same slope. At an intermediate time, the particle numbers of the two halves equal, the curves cross each other, oscillating behaviour starts, with decreasing oscillation amplitude, and for a very long time the particle numbers will be equal. The slope of the initial decrease/increase is independent of the system size, and for $t \ll L$ the increasing curves are identical.   
	
	In Fig. \ref{fig:Nvsh}  the particle numbers are presented after different values of $h$, and for a given system size ($L=144$). All of these quenches are in the extended phase,  most of them close to the critical point. As the quench is closer and closer to the critical point, the dynamics (slope of the curves) become slower and slower. The first crossing of the growing and decreasing particle numbers occurs later and later.
	I call the time of the first crossing  "crossing time". One can obtain these crossing times using Fig. \ref{fig:NCRIT}, and one can define a time scale using the crossing time.
	
	The crossing times are presented in Fig. \ref{fig:crossingtimes} as the function of the $h$ parameter.  I have found, that the crossing time diverges approaching the critical point. The numerical fit shows $0.98\pm0.03$ as the critical exponent, therefore the scaling of the first crossing time is likely to be
	\begin{eqnarray}
		t_{\textrm{cross}} \sim \frac{1}{|1-h|} \;.
	\end{eqnarray} 
	
	In Fig. \ref{fig:NCRIT} the (log of the) growing particle number of the initially empty subsystem is shown as the function of $\ln t$.  The particle number approximately grows as
	\begin{equation}
		N_B \sim t^{0.57} \;,
	\end{equation} 
	nut there are strong oscillations on the overall trend, which makes the measurement of the exponent difficult. The exponent obtained here for the particle number is bigger than the exponents of the entanglement entropy and magnetization obtained in \cite{roosz2014} for the case of the global quenches.  
	
	In the context of front dynamics, it is usual to investigate the shape of the front. Here the local filling $n_l = \langle c^{\dagger}_l(t) c_l(t) \rangle$ is investigated. However, the local value of this filling strongly depends on the on-site potential, and if one simply calculates it, no trend can be seen, only rapid oscillations.
	To define a meaningful front shape, one introduces a $\varphi$ phase to the Hamiltonian 
	\begin{equation}
		H=-\frac{1}{2}\sum_{l=1}^{L-1} c^{\dagger}_l c_{l+1} + c^{\dagger}_{l+1} c_l  + h \sum_{l=1}^L \cos\left( 2 \pi \kappa l  + \varphi \right) c^{\dagger}_l c_l \;,
		\label{eq:ham_phase} 
	\end{equation}
	and averages over the  $\varphi$ phase. Different $\varphi$ phases mean different relative positions of the lattice and the potential. So averaging over the phase is equivalent to averaging over the position where the boundary of the initially distinct two subsystems is located. In numerical calculations $10^4$ randomly chosen values of $\varphi$ have been used (from the uniform distribution on $[0,2 \pi]$).

	The scaled front shapes are shown in Fig. \ref{fig:scaling}. Here the particle numbers are averaged over $10^4$ randomly chosen phases in order to get relatively smooth results. The front shapes can be scaled together, and the scaling exponent $\alpha$ is different in the extended phase, at the critical point, and in the localized phase.   The scaling function $\Phi(x)$ (which is also different in the aforementioned tree cases) can be defined as follows
	\begin{equation}
		\Phi\left( \frac{j-L/2}{t^{\alpha}} \right)=n_{j} \;.
	\end{equation} 
	In the localized phase, the scaling is trivial, after a short initial evolution the front reaches a constant shape "froze in" - this corresponds to $\alpha_{\textrm{localized}}=0$.  In the extended phase, the scaling exponent is found to be one   $\alpha_{\textrm{extended}}=1.0$, which is identical to the scaling of the homogeneous XX chain front \cite{hunyadi2004}. In the critical point, the best fit has been found at 
	$\alpha_{\textrm{critical}}=0.55$
	
	
	I also investigated the mutual information in the zero temperature quench between neighboring intervals at the initial connection point ( one interval is from $L/2-l$ to $L/2$, and the other one is from $L/2+1$ to $L/2+l$). One expects, that the initial growth of the mutual information and the entanglement entropy are closely related, and the mutual information has more information about the correlations in the system. The numerical results are shown in \ref{fig:zero-T-MUTI-size-dep}, \ref{fig:zero-T-MUTI-size-dep-log}, \ref{fig:scaling}. In figure    \ref{fig:zero-T-MUTI-size-dep} the mutual information is shown as a function of the subsystem size $l$ in a given time $t=20$. There is an initial growing region, and then a constant plateau starts. In figure   \ref{fig:zero-T-MUTI-size-dep}  the mutual information is plotted as a function of the logarithm of the subsystem size.  For the extended phase the mutual information for very small systems changes very slowly, for intermediate sizes there is a logarithmic growth, and then the plateau starts. Interestingly, at the critical point, there is no very slow initial region, and the mutual information is proportional to the logarithm of the subsystem size from very small sizes.
	The numerical data is compatible with
	\begin{equation}
		I_{T=0, extended} \sim 1.1 \ln l \;,
	\end{equation}  
	\begin{equation}
		I_{T=0, critical} \sim 0.51 \ln l \;.
	\end{equation}
	My results about the mutual information are in agreement with the results of \cite{roy2021}, they calculated the entanglement entropy, not the mutual information however for short times when the width of the front is smaller than the subsystem size   $l$ the time dependence of these two quantities has to be the same, and the numerical results are indeed very similar.
	
	Results about the time dependence of the mutual information for fixed system sizes are shown in Fig. \ref{fig:zero-T-MUTI-time-dep}. Here one finds, that the mutual information is proportional to the logarithm of the time, and the prefactors differ in the critical point and in the extended phase. In the localized phase the mutual information converges to a constant value, independent of the subsystem and system sizes. The numerical data suggest 
	\begin{equation}
		I_{T=0, extended} \sim 1.15 \ln t \;,
	\end{equation}  
	\begin{equation}
		I_{T=0, critical} \sim 0.55 \ln t \;.
	\end{equation}
	I calculated the mutual information in the finite temperature quench. The temperatures have been $T_A=0.5$ and $T_B=2.0$.


	
	In Fig. \ref{fig:fin-T-ENERGY} the time dependence of the energy of the two subsystems is shown. The initial temperatures were $T_A=1$ and $T_B=10$. In the extended phase, and at the critical point the energy values approaching each other, the initially bigger value decrease and the initially smaller value increase. However, in the localized phase, the energy values remain almost constant. 
	
	In Fig. \ref{fig:fin-T-MUTI} the mutual information is shown after a finite temperature quench without averaging over the phase, while one can conclude, that in the localized phase the mutual information remains bounded, and it grows in the extended phase and at the critical point, the oscillations are strong, and it is hard to find the overall trend. 
	
	In Fig. \ref{fig:fin-T-MUTI-time-dep-h0.5} the mutual information is shown after a finite temperature quench averaged over 100 randomly chosen phases. There is an initial region (while the front size does not reach the system size), where the mutual information grows with the logarithm of the time. This initial growth is found to be
	\begin{equation}
		I_{T>0,extended} = 0.2 \ln (t) - 0.25 \;.
	\end{equation}  
	At the critical point (see \ref{fig:fin-T-MUTI-time-dep-h0.5} ) the initial growth of the mutual information is 
	\begin{equation}
		I_{T>0,critical} = 0.12 \ln (t) - 0.07 \;.
	\end{equation}  
	In Fig. \ref{fig:fin-T-MUTI-size-dep} the mutual information is shown as a function of the subsystem size. There is an initial slow region, a middle region, where the variation of the mutual information is proportional to the logarithm of the subsystem size, and for big subsystem sizes the mutual information converges to a constant value. The  middle logarithmic part is found to be 
	\begin{equation}
		I_{T>0, extended} = 0.48 \ln l  -1.2
	\end{equation}	
	\begin{eqnarray}
		I_{T>0, critical} = 0.21 \ln l - 0.14
	\end{eqnarray}
	\section{Conclusions}
	Here the front dynamics of a non-interacting model with a localisation transition have been investigated. In the extended phase the dynamics are qualitatively similar to the homogeneous XX front dynamics \cite{viktor2014}. However, when approaching the critical point, the time scale of the dynamics (defined by the crossing time in this work) diverges with exponent one, and in the critical point slower, diffusive dynamics occur.
	In the case of the zero-temperature quench, where the initial difference is the particle number, the front shapes can be scaled together. For the localized phase the scaling is trivial, for the extended phase, the scaling is equivalent to the scaling of the homogeneous system. In the critical point, the scaling includes a power law $t^{0.5}$. This exponent is close to the literature value of the wave packet spreading exponent of the Harper model \cite{wilkinson1994,roosz2020} which is known to be $0.477$. There is a simple reasoning behind this phenomenon, if half of a system is filled, the other half empty, one can perform a particle-hole transformation on the filled half of the system, and in the resulting effective system, the problem is equivalent to a local quench. However, the initial state used here is not fully empty/fully filled, but one half is initially empty, and the other half is initially half filled - this may cause the difference between the wave packet scaling exponent known from the literature and the front shape scaling exponent measured here. A similar simple connection between the finite temperature initial state and the local quenches does not exist.        
	The scaling of the mutual information with the subsystem size and the time is found to be logarithmic bought in the extended phase and in the critical point. The prefactors of the time and subsystem size dependence agree up to the precision of this work in the zero temperature quench, however 
	in the finite temperature quench they are significantly different.
	
	This is a major difference compared to the homogeneous XX chain, where the prefactor of  the logarithmic scaling of the   mutual information is the same for the time and space dependence. 
	
	The prefactors of the mutual information in the critical point are generally smaller than in the extended phase, this phenomenon is similar to the entanglement scaling found in \cite{roosz2020}.

	Further generalizations of the present work may include numerical studies about the dependence of the dynamics on the irrational potential, or numerical studies in a generalization of the present model where the quasiperiodic potential is $\cos( 2 \pi \kappa l^{\alpha} )$, in the details of the transition depends on the $\alpha$ exponent \cite{imre1992}.

	\section{ACKNOWLEDGMENTS}
	The author thanks F. Igl\'oi, R. Juh\'asz,  and
	Z. Zimbor\'as for useful discussions. This work was supported 
	by the National Research, Development and Innovation
	Office NKFIH under Grant No. K128989. This work was supported
	by the National Quantum Information Laboratory of Hungary. 
	
	\appendix
	\section{Time evolution}
	I use the Heisenberg picture. It is possible to threaten the two protocols in a unified way. The only difference is the calculation of the initial correlations. 
	The initial state is characterized by its two-point correlation function
	\begin{eqnarray}
		g_{l,m} = \langle c^{\dagger}_l c_m\rangle
	\end{eqnarray}  
	In the case of the first protocol (particle number quench) to calculate this correlation function I used a   modified initial Hamiltonian. I added a big ($100$)  local potential to one ($A$) part of the Hamiltonian and diagonalized this modified Hamiltonian. With this high potential, the $A$ part is practically empty.
	\begin{align}
		H&=-\frac{1}{2}\sum_{l=1}^{L-1} c^{\dagger}_l c_{l+1} + c^{\dagger}_{l+1} c_l  + h \sum_{l=1}^L \cos\left( 2 \pi \kappa  l \right) c^{\dagger}_l c_l \\
		&+ 100 \sum_{l=1}^{L/2} c_l^{\dagger} c_l
		\label{eq:ham_mod}
	\end{align}
	One diagonalises the modified Hamiltonian with a canonical transformation 
	\begin{align}
		\eta_k&=\sum_{l=1}^L u^{(0)}_{k,l} c_l\\
		H&=\sum_{k=1}^L \epsilon^{(0)}_{k} \eta^{\dagger}_k \eta_k
	\end{align} 
	and the initial correlations are given by
	\begin{equation}
		g_{l,m}=\sum_{k, \epsilon_k<0} u_{k,l} u_{k,m}
	\end{equation}
	For the second protocol (temperature quench) one diagonalises $H_A$ and $H_B$ separately, the correlations in $A$ ($B$) are determined by $H_A$ ($H_B$) and the correlations between the two subsystems are zero.
	\begin{align}
		H_A&=-\frac{1}{2}\sum_{l=1}^{L/2-1} c^{\dagger}_l c_{l+1} + c^{\dagger}_{l+1} c_l  + h \sum_{l=1}^{L/2} \cos\left( 2 \pi \kappa  l \right) c^{\dagger}_l c_l \\
		H_B&=-\frac{1}{2}\sum_{l=L/2}^{L-1} c^{\dagger}_l c_{l+1} + c^{\dagger}_{l+1} c_l  + h \sum_{l=L/2}^{L} \cos\left( 2 \pi \kappa  l \right) c^{\dagger}_l c_l \\
		\label{eq:H_A_H_B}
	\end{align}
	\begin{align}
		\eta^{A}_k&=\sum_{l=1}^{L/2} u^{A,(0)}_{k,l} c_l\\
		H&=\sum_{k=1}^{L/2} \epsilon^{A,(0)}_{k} \eta^{\dagger}_k \eta_k
	\end{align}
	\begin{align}
		\eta^{B}_k&=\sum_{l=L/2}^{L} u^{B,(0)}_{k,l} c_l\\
		H&=\sum_{k=L/2}^L \epsilon^{B,(0)}_{k} \eta^{\dagger}_k \eta_k
	\end{align} 
	The correlation function is given by 
	\begin{align}
		g_{l,m}&=\\
		& \left\{ \begin{array}{c}
			\sum_{k} u^{A,(0)}_{k,l} u^{A,(0)}_{k,m} n(\epsilon_k, T_A) \quad \textrm{if} \quad l,m \in A \\
			\sum_{k} u^{B,(0)}_{k,l} u^{B,(0)}_{k,m} n(\epsilon_k, T_B) \quad \textrm{if} \quad l,m \in B \\
			0 \quad \textrm{if} \quad l \in A \,\textrm{and}\, m \in B \\
			0 \quad \textrm{if} \quad l \in B \,\textrm{and}\, m \in A
		\end{array}  \right. \nonumber
	\end{align}
	where $n(\epsilon, T)=1/(\exp[\epsilon/T]+1))$ is the Fermi function.
	
	For $t>0$ one diagonalizes the after quench Hamiltonian 
	\begin{align}
		\nu_k&=\sum_{l=1}^{L} u_{k,l} c_l\\
		H&=\sum_{k=1}^L \epsilon_{k} \nu^{\dagger}_k \nu_k
	\end{align} 
	The time evolution of the $\nu$ operators in the Heisenberg picture is $\nu(t)=e^{-i \epsilon_k t} \nu(0)$,  $\nu^{\dagger}(t)=e^{-i \epsilon_k t} \nu^{\dagger}(0)$. 
	
	The time evolution of  the $c_l$ operators is
	\begin{align}
		c_l(t) &= \sum_{k=1}^L u_{k,l} e^{-i \epsilon_k t} \nu(0) \\ 
		&= \sum_{l=1}^L \sum_{m=1}^L  \sum_{k=1}^L u_{k,l} e^{-i \epsilon_k t} u_{k,m} c_{m}(0) \\
		&= \sum_{l=1}^L \sum_{m=1}^L f(m,l,t) c_m(0) \\
		f(m,l,t) &= \sum_{k=1}^L u_{k,l} e^{-i \epsilon_k t} u_{k,m}
		\label{eq:time_ev}
	\end{align}
	The time evolution of the particle number operators $\hat{N}_A$, $\hat{N}_B$  and the local Hamiltonians $H_A$, $H_B$ can be written as bilinear expressions using E.q. (\ref{eq:time_ev}). The expecation values are
	\begin{align}
		\langle \hat{N}_A \rangle  &= \sum_{l=1}^{L/2} \sum_{n=1}^L \sum_{m=1}^L  f^*(n,l,t) f(m,l,t)  g_{n,m} \\
		\langle \hat{N}_B \rangle  &= \sum_{l=L/2}^{L} \sum_{n=1}^L \sum_{m=1}^L  f^*(n,l,t) f(m,l,t)  g_{n,m} \\
		\langle H_A \rangle  &=\\
		&\sum_{l=1}^{L/2} \sum_{n=1}^L \sum_{m=1}^{L/2}  f^*(n,l,t) f(m,l,t)  g_{n,m} h \cos\left( 2 \pi \kappa  l \right) \nonumber \\
		&+\sum_{n=1}^{L}\sum_{m=1}^{L} \frac{1}{2} (f^*(n,l+1,t) f(m,l,t)+ \nonumber \\
		&\  f(n,l+1,t) f^*(m,l,t))  g_{n,m} \nonumber \\
		\langle H_B \rangle  &= \\
		&\sum_{l=L/2}^{L} \sum_{n=1}^L \sum_{m=1}^L  f^*(n,l,t) f(m,l,t)  g_{n,m}   h \cos\left( 2 \pi \kappa  l \right) \nonumber \\
		&+\sum_{n=1}^L \sum_{m=1}^{L} \frac{1}{2} (f^*(n,l+1,t) f(m,l,t) \nonumber \\
		&+ f(n,l+1,t) f^*(m,l,t))  g_{n,m} \nonumber
	\end{align}


\begin{thebibliography}{100}

\bibitem{antal1999} T. Antal, Z. Rácz, A. Rákos, and G. M. Schütz, Phys. Rev. E
		59, 4912 (1999).

\bibitem{ogata2002} Y. Ogata, Phys. Rev. E 66, 066123 (2002).

\bibitem{karevski2002} D. Karevski, Eur. Phys. J. B 27, 147 (2002)

\bibitem{hunyadi2004} V. Hunyadi, Z. Rácz, and L. Sasvári
		PHYSICAL REVIEW E 69, 066103 (2004)

\bibitem{antal1997} T. Antal, Z. R\'acz, and L. Sasv\'ari, Phys. Rev. Lett. 78, 167
		(1997)

\bibitem{antal1998} T. Antal, Z. R\'acz, A. R\'akos, and G. M. Sch\"utz, Phys. Rev. E
		57, 5184 (1998)

\bibitem{viktor2014} Viktor Eisler, Zoltan Zimboras Phys. Rev. A 89, 032321 (2014)

\bibitem{viktor2013} Viktor Eisler and Zoltán Rácz Phys. Rev. Lett. 110, 060602 (2013)

\bibitem{platini2007} 
		Thierry Platini and Dragi Karevski J. Phys. A: Math. Theor. 40 1711 (2007)

\bibitem{hiroki2019} 
		Hiroki Moriya, Rikuo Nagao, and Tomohiro Sasamoto
		J. Stat. Mech. 2019 063105 (2019)

\bibitem{federico2022} 
		Federico Rottoli, Stefano Scopa, and Pasquale Calabrese
		J. Stat. Mech. 2022 063103 (2022)

\bibitem{marton2017} 
		M\'arton Kormos SciPost Phys. 3 020 (2017)

\bibitem{gambassi2017} 
		Gabriele Perfetto and Andrea Gambassi
		Phys. Rev. E 96 012138 (2017)

\bibitem{dasgupta2017} 
		Sirshendu Bhattacharyya and Subinay Dasgupta
		Eur. Phys. J. B 90 140 (2017)

\bibitem{viktor2016} 
		Viktor Eisler, Florian Maislinger, and Hans Gerd Evertz
		SciPost Phys. 1 014 (2016)

\bibitem{marton2017B} 
		Márton Kormos, Zoltán Zimborás  J. Phys. A: Math. Theor. 50 264005 (2017)

\bibitem{platini2006} 
		T. Platini and D Karevski J. Phys.: Conf. Ser. 40 93 (2006)

\bibitem{platini2005} Platini, T., Karevski, D. Scaling and front dynamics in Ising quantum chains. Eur. Phys. J. B 48, 225-231 (2005)

\bibitem{viktor2020} 
		Viktor Eisler and Florian Maislinger
		SciPost Phys. 8 037 (2020) 
		
		%

\bibitem{scopa2022} 
		Stefano Scopa , Pasquale Calabrese, Jérôme Dubail
		SciPost Phys. 12, 207 (2022) (2022)

\bibitem{lopez2021} 
		Javier Lopez-Piqueres, Brayden Ware, Sarang Gopalakrishnan, and Romain Vasseur
		Phys. Rev. B 104 104307 (2021)

\bibitem{collura2020} 
		Mario Collura, Andrea De Luca, Pasquale Calabrese, and Jérôme Dubail
		Phys. Rev. B 102, 180409(R) (2020)

\bibitem{vir2019} 
		Vir B. Bulchandani and Christoph Karrasch
		Phys. Rev. B 99 121410 (2019)

\bibitem{braunfield2017} 
		Viktor Eisler and Daniel Bauernfeind
		Phys. Rev. B 96 174301 (2017)

\bibitem{misguich2013} Thiago Sabetta and Grégoire Misguich
		Phys. Rev. B 88 245114 (2013)

\bibitem{einhellinger2012} M. Einhellinger, A. Cojuhovschi, and E. Jeckelmann
		Phys. Rev. B 85, 235141 (2012)

\bibitem{lancester2010} 
		Jarrett Lancaster and Aditi Mitra
		Phys. Rev. E 81, 061134 (2010)

\bibitem{langer2009} 
		S. Langer, F. Heidrich-Meisner, J. Gemmer, I. P. McCulloch, and U. Schollwöck
		Phys. Rev. B 79, 214409 (2009)

\bibitem{antal2008} 
		T. Antal, P. L. Krapivsky, and A. Rákos
		Phys. Rev. E 78, 061115 (2008)

\bibitem{dominique2005} 
		Dominique Gobert, Corinna Kollath, Ulrich Schollwö?ck, and Gunter Schütz
		Phys. Rev. E 71, 036102 (2005)

\bibitem{roosz2014} G Roósz, U Divakaran, H Rieger, F Iglói
		Physical Review B 90 (18), 184202 (2014)

\bibitem{lieb61} E. Lieb, T. Schultz, and D. Mattis, Ann. Phys. (N.Y.) 16,
		407 (1961)

\bibitem{jordan1928} Jordan P and Wigner E 1928 Z. Phys. 47 631

\bibitem{zimbi2014} 
		Viktor Eisler, Zoltan Zimboras New J. Phys. 16 (2014) 123020

\bibitem{XHorvath2022} 
		Dávid X. Horváth, Spyros Sotiriadis, Márton Kormos, Gábor Takács
		SciPost Phys. 12, 144 (2022)

\bibitem{calabrese2007} P. Calabrese and J. Cardy, J. Stat. Mech. (2007) P06008

\bibitem{calabrese2005} P. Calabrese and J. Cardy, J. Stat. Mech. P04010 (2005)

\bibitem{igloi2011} F. Igl\'oi and H. Rieger, Phys. Rev. Lett. 106, 035701
		(2011).

\bibitem{rieger2011} H. Rieger and F. Igl\'oi, Phys. Rev. B 84, 165117 (2011)

\bibitem{alba2019} V. Alba, B. Bertini and M. Fagotti, SciPost Phys. 7, 005 (2019)

\bibitem{bertini2018} B. Bertini, M. Fagotti, L. Piroli and  P. Calabrese J. Phys. A: Math. Theor. 51, 39LT01

\bibitem{castro2016} O. A. Castro-Alvaredo, B. Doyon and T. Yoshimura, Phys. Rev. X 6, 041065 (2016)

\bibitem{ruggiero2020} P. Ruggiero, P. Calabrese, B. Doyon and J. Dubail, 
		Phys. Rev. Lett. 124, 140603 (2020)

\bibitem{doyon2020} B. Doyon,  SciPost Phys. Lect. Notes 18, (2020)

\bibitem{misguich2019} G. Misguich, N. Pavloff and V. Pasquier, SciPost Phys. 7, 025 (2019)

\bibitem{alba2021} V. Alba, B. Bertini, M. Fagotti, L. Piroli and P. Ruggiero, J. Stat. Mech. 114004 (2021)

\bibitem{dubail2022} I. Bouchoule and J. Dubail, J. Stat. Mech. 014003 (2022)

\bibitem{fagotti2017} M. Fagotti, Phys. Rev. B 96, 220302 (2017)

\bibitem{ruggiero2021} P. Ruggiero, P. Calabrese, B. Doyon and J. Dubail, J. Phys. A: Math. Theor. 55, 024003 (2021)

\bibitem{bastellino2019} A. Bastianello, V. Alba and J.-S. Caux, Phys. Rev. Lett. 123, 130602 (2019)

\bibitem{bulchandani2017} V. B. Bulchandani, R. Vasseur, C. Karrasch and J. E. Moore, Solvable hydrodynamics of quantum integrable systems, Phys. Rev. Lett. 119, 220604 (2017)

\bibitem{doyon2018} B. Doyon, T. Yoshimura and J.-S. Caux, Phys. Rev. Lett. 120, 045301 (2018)

\bibitem{schemmer2019} M. Schemmer, I. Bouchoule, B. Doyon and J. Dubail,  Phys. Rev. Lett. 122, 090601 (2019)

\bibitem{malvania2021} N. Malvania, Y. Zhang, Y. Le, J. Dubail, M. Rigol and D. S. Weiss, Science 373, 1129 (2021)

\bibitem{igloi2012} F Iglói, Z Szatmári, YC Lin
		Physical Review B 85 (9), 094417 (2012)

\bibitem{igloi2013} F Igl\'oi, G Ro\'osz, YC Lin, New Journal of Physics 15 (2), 023036 (2013)

\bibitem{roosz2017} G Ro\'osz, YC Lin, F Igl\'oi
		New Journal of Physics 19 (2), 023055 (2017)

\bibitem{aubry1980} S. Aubry and G. Andre, Ann. Israel Phys. Soc. 3 133
		(1980).

\bibitem{thouless1983} D. J. Thouless, Phys. Rev.B 28, 4272 (1983)

\bibitem{harper1955} P. G. Harper, Proc. Phys. Soc. A 68, 874 (1955).

\bibitem{evers2008} F. Evers and A. D. Mirlin, Anderson transitions, Rev.
		Mod. Phys. 80, 1355 (2008).

\bibitem{tessa2020} 
		Tessa Cookmeyer, Johannes Motruk, and Joel E. Moore
		Phys. Rev. B 101, 174203 (2020)

\bibitem{sinha2019} 
		Aritra Sinha, Marek M. Rams, and Jacek Dziarmaga Phys. Rev. B 99, 094203 (2019)

\bibitem{wilkinson1994} M. Wilkinson and E. J. Austin, Phys. Rev. B 50, 1420 (1994).

\bibitem{evang200} S. N. Evangelou and J.-L. Pichard, Phys. Rev. Lett. 84, 1643
		(2000).

\bibitem{siebesma87} A. P. Siebesma and L. Pietronero, Europhys. Lett. 4, 597 (1987).

\bibitem{roosz2020} G. Ro\'osz, Z Zimbor\'as, R Juh\'asz
		Physical Review B 102 (6), 064204  (2020)

\bibitem{roy2019} Nilanjan Roy, Auditya Sharma Phys. Rev.  B 100, 195143 (2019)

\bibitem{hauschild2015} 
		Johannes Hauschild, Frank Pollmann, and Fabian Heidrich-Meisner
		Phys. Rev. A 92 053629 (2015)

\bibitem{rozenheimer2013} J. P. Ronzheimer, M. Schreiber, S. Braun, S. S. Hodgman, S. Langer, I. P. McCulloch, F. Heidrich-Meisner, I. Bloch, and U. Schneider
		Phys. Rev. Lett. 110, 205301 (2013)

\bibitem{modungo2009} M. Modugno, New J. Phys. 11 033023 (2009)

\bibitem{deisler2011} B. Deissler, E. Lucioni, M. Modugno, G. Roati, L. Tanzi, M. Zaccanti, M. Inguscio and G. Modugno, New J. Phys. 13, 023020 (2011).

\bibitem{choi2016} Jae-yoon Choi, Sebastian Hild, Johannes Zeiher, Peter Schauß, Antonio Rubio-Abadal, Tarik Yefsah, Vedika Khemani, David A. Huse, Immanuel Bloch, Christian Gross; Science 352, 1547 (2016)

\bibitem{bennett1996} C. H. Bennett, H. J. Bernstein, S. Popescu, and B. Schumacher, Phys. Rev. A 53, 2046 (1996)

\bibitem{nielsen2020} M.A. Nielsen and I.L. Chuang, Quantum Computation
		and Quantum Information (Cambridge University Press,
		Cambridge, 2000).

\bibitem{roy2021} 
		Aamna Ahmed, Nilanjan Roy, and Auditya Sharma
		Phys. Rev. B 104, 155137 (2021)

\bibitem{imre1992} I Varga, J Pipek, B Vasvári
		Physical Review B 46 (8), 4978 (1992)


\end{thebibliography}
\end{document}